\documentstyle[aps,preprint]{revtex}
\newcommand{\C}{1\!\!\!C}

\newcommand{\R}{I\!\!R}
\newcommand{\Sc}{{\cal S}}
\newcommand{\be}{\begin{eqnarray*}}
\newcommand{\ee}{\end{eqnarray*}}

\begin{document}
\draft
\title{
Complex structure and the construction of\\
 the  $:\!\phi^4_4\!:$ quantum field theory
  in four-dimensional space-time
\thanks{This work
is supported partly by
Russian Fundamental Research Foundation under Grant
No.  96-01649.  It is  the announce  paper of the project
$\phi^4_4\cap M.$ } }

\author{Edward  P. Osipov}
\address{Department of Theoretical Physics,
 Sobolev Institute for Mathematics,
\\  630090 Novosibirsk, RUSSIA
\\ E-mail address: osipov@math.nsk.su}

\maketitle
\medskip

\begin{abstract}
We announce results about the nonperturbative
mathematically rigorous construction of the
 $:\!\phi^4_4\!:$
  quantum field theory
   in four-dimensional space - time.
The complex structure of solutions
of the classical nonlinear (real-valued) wave equation and
quantization are closely connected among themselves and allow to
construct non-perturbatively the quantum field theory  with
interaction $:\!\phi^4_4\!:$ in four-dimensional  space - time.
We consider vacuum averages, in particular, we construct
Wightman functions and matrix elements of the scattering operator
as  generalized functions for finite energies.  The constructed
theory is obviously  nontrivial.

\end{abstract}
\medskip
\medskip


\medskip
\medskip
In this Letter we announce a part of results about
the nonperturbative mathematically rigorous construction of the
 $:\!\phi^4_4\!:$
  quantum field theory
   in four-dimensional space - time.
   The announced results are steps in realization of
the project $\phi^4_4\cap M$.

The construction of the quantum field is based on properties of
dynamics of the classical system.
Namely, there exists an intimate relationship between the dynamics
of classical system, its complete integrability, its
 complex structure, and quantization.
 About  the
  complete integrability of
	the classical  $u^4_4$ interaction, see
	for  instance,
 \cite{Bae92}.
The complex structure
for the classical nonlinear wave equation
with interaction $u^4_4$
  was constructed  in
 \cite{Osi95a},
 see also
 \cite{Osi96a}.
 We formulate the statement  about the
complex  structure in a convenient form.

We consider the solutions of the
classical nonlinear wave equation
   in four-dimensional space - time
$$
{ \partial^2\over\partial t^2} u
-\Delta u + m^2 u + \lambda u^3=0,\quad m>
0,\quad\lambda > 0.\eqno (1) $$

A solution is given uniquely by its  initial data
(by its canonical coordinate and canonical
momentum), for instance,
at time zero,
 $$ \varphi (x) =u (t, x) |_{t=0},
 \quad \pi (x) = {\partial\over\partial t} u (t, x) |_{t=0},
 $$
 and these initial data belong to the space
  $ H^1\oplus L_2,$ $(\varphi,\pi) \in H^1\oplus L_2.$
 Let $U (t), $  $W,$ and $S$ be, respectively,
  the operator of dynamics, the wave operator to the
 backward, and  the scattering operator of the
 nonlinear wave equation (1).
  These
nonlinear operators are correctly defined for
 initial data from $H^1\oplus L_2$, are invertible
and are canonical transforms (i.e. symplectomorphisms)
 (see, for instance, \cite{MorS72a,MorS72b,Bae92}).

Let $R$ be the map
$$ R(\varphi,\pi) =\varphi + i\mu^ {-1}
\pi\equiv\varphi^+, \quad
R^{-1}\varphi^+ = (\mbox {Re}\, \varphi^+,
\mu\, \mbox{Im}\, \varphi^+), $$
where $\mu= (-\Delta + m^2)^{1/2} $.
$R$ maps an initial data on the positive frequency part
 of the free solution with this initial data (at time zero).

The essential feature of the construction of
the  $:\!\phi^4_4\!:$
quantum field theory is the
statement  about the complex  structure of the
 classical nonlinear wave equation (1).

\medskip
{\bf Theorem 1 (complex structure)}
(see
 \cite[Theorem 1.1]{Osi95a}).
{\it The maps  $RU(t)R^{-1},$ $RWR^{-1},$ $RSR^{-1}$
are correctly defined
and are complex holomorphic maps on the
space $H^1(\R^3,\C)$ into itself.
In particular,  for $$ z(\alpha)
=\sum^N_{j=1} \alpha_j z_j,\quad z_j\in H^1(\R^3,\C),\quad
\alpha_j\in\C,\quad h\in\Sc (\R^3,\C), $$
the functions
$ \int d^3x (RU(t)R^{-1}z(\alpha))(x)h(x),$
 $$\int d^3x (RWR^{-1}z(\alpha))(x)h(x),
 \quad \int d^3x (RSR^ {-1} z(\alpha))(x)h(x) $$
are complex holomorphic functions
 on $ (\alpha_1,...,\alpha_N) \in \C^N.$
}
\medskip

The construction of the quantum field can
be done by several different
ways (which are in agreement with themselves).
Here we consider the one of them.

To  construct  the
  $:\!\phi^4_4\!:$
quantum field theory we use the existence of the
wave operator of the classical nonlinear equation (1) and
the statement  of Theorem 1 about the  complex structure.
 We
can define the interacting  field as
$$ \phi= {1\over 2}
(:\! RWR^ {-1} (\phi_ {in}) \!:
+ :\! R\Theta^T W\Theta^T R^ {-1} (\phi_
{in}) \!:,\eqno (2) $$
where
$\Theta^T$ is the operator
of  time reflection  for
  the  classical
 system,
$\phi_ {in} $ is the incoming
quantum field (it is free), $:\!.\!:$ means the Wick normal
ordering with respect to the free quantum field $\phi_ {in}.$
 Expression (2) gives the possibility to define
 the Wick kernel of the
interacting quantum field
( the Wick kernel is equal to matrix elements
 of   (2) on coherent vectors).
 We note that on the  diagonal the Wick
 symbol of the interacting quantum field (2) is a (real)
  solution of the classical nonlinear equation (1),
  and it is namely the same solution, in which
the wave operator maps the free solution
given by the diagonal of the Wick symbol of the
free quantum field $\phi_ {in}. $

	Taking into account
the holomorphity of the map $RWR^ {-1}$
	and the $T$-symmetry of Eq. (1)
the expression (2) has quite correct sense.
In particular, it  is a bilinear form defined on
  coherent vectors of the Fock space
of the $in$-field  (see
\cite{Hei74,Rac75,Osi94a,Osi95a,Osi96b}).

The other way of construction of the
quantum field (and the construction of its Wick kernel)
is the  definition of nonlinear quantum equation
 in the integral
form,
$$ \phi (t, x) =\phi_ {in} (t, x) -\lambda\int^t_ {-\infty}
\int d\tau d^3y R(t-\tau, x-y):\!\phi^3 (\tau, y) \!:
\eqno (3) $$
with  normal ordering with respect
to the $in$-field,
see \cite{Hei74,Osi94a,Osi94b,Rac75}.
This way is consistent with the definition of
 quantum field given by (2).

The introduction of the quantum field in the form
(2), or the construction of the bilinear form
--  solution of Eq.  (3), allows to write out
 the explicit expression for
the total Hamiltonian. The total Hamiltonian
 is the self-adjoint  positive operator.
 It is correctly  defined
 as a bilinear form on coherent
  vectors from $D_{coh} (H^1 (\R^3,\C)). $
 The total Hamiltonian
 is equal to the following expressions
 \be
H&=&{1\over 2}\int d^3 x\left(:\!\dot\phi^2(t,x)\!:+
:\!\nabla\phi^2(t,x)\!:+m^2:\!\phi^2(t,x)\!:+{\lambda\over 2}
 :\!\phi^4(t,x)\!:\right)\cr &=&{1\over 2}\int d^3
x\left(:\!\dot\phi^2_{in}(t,x)\!:+
:\!\nabla\phi^2_{in}(t,x)\!:+m^2:\!\phi^2_{in}(t,x)\!:\right)\cr
&=&{1\over 2}\int d^3 x\left(:\!\dot\phi^2_{out}(t,x)\!:+
:\!\nabla\phi^2_{out}(t,x)\!:
+ m^2:\!\phi^2_{out}(t,x)\!:\right),
 \ee
see
\cite{Hei74,Osi94a,Osi95a,Osi96b}.

The possibility to construct the Wick kernel, or
the Wick symbol,  of the interacting quantum field
simplifies a consideration of its properties
and the construction of vacuum averages.

The consideration of this
Wick kernel and the construction
of bilinear form is based on
  the Fock-\-Bargmann-\-Berezin-\-Segal
 integral representation
  over  coherent states,
	see \cite{Per86,PanPSZ91,KlaS85}.
For this description we use
initial data of the $in$-field. This
 choice of coordinates
 is natural from the
physical point of view.
 It  requires to  consider integrals
 with Gaussian promeasure and
 simplifies  estimates
and the construction of
 operator--\-valued generalized function,
corresponding to the Wick kernel of the interacting field,
see
\cite{Osi96b,Osi96c}.
 Here we do not present the exposition
of these results important for the formulation
of locality condition, we only illustrate
them by the consideration of
smoothed quantum field for
finite  momentum and finite energy.

The consideration of finite  energies
(for our massive case!) is technically
 more simple and allows to
 construct  vacuum averages.

To construct  vacuum averages
we  consider
 the following smoothing of
the quantum field
$$ \int dt_1 dt_2 d^3x_1
d^3x_2\,e^{iHt_1+iPx_1}\,\phi(0,0)\,e^{iHt_2+iPx_2}\,
f_1(t_1,x_1)f_2(t_2,x_2).\eqno(4) $$
 The expression
$\phi(t,x)\vert_{t=0,x=0}$
 is correctly defined as a bilinear form
 and (4) corresponds to the following
 bilinear form
 $$ f^\sim_1(H,P)\phi(0,0)f^\sim_2(H,P), $$
 where
$f^\sim_1, f^\sim_2$ is the Fourier transform
 of functions $f_1,$ and, respectively,
 $f_2,$  $H$ is
    the total Hamiltonian and $P$ is the
    operator  of momentum.
    As test functions we shall take
    smooth    functions
with compact support in momentum space.
	Using  holomorphity (Theorem 1)
it is possible to write the Taylor expansion (at zero)
 $$
\phi=\sum^\infty_{n=1}\,
\phi_n :\!\underbrace{\phi_{in}...\phi_{in}}_{n}\!:,
$$
coefficients
 $\phi_n$ of this expansion
 are equal to the corresponding derivatives of the
 classical wave operator,
$$
\phi_n={1\over 2}n!^{-1}(d^n RWR^{-1}(0) +
d^n R\Theta^T W\Theta^T R^{-1}(0)).
$$
These Taylor coefficients $\phi_n$ are correctly
defined as Schwartz distributions,
$\phi_n\in\Sc'(\R^{3n})$.

Since we use as test functions smooth functions with
 compact support  in momentum space
 and, taking into account, that we consider
the massive case, it is easy to see, that
 $$ f^\sim_1(H,P)\phi(0,0)f^\sim_2(H,P)
 = \sum^{N(f_1,f_2)}_1
f^\sim_1(H,P)\phi_n(:\!\underbrace{\phi_{in}...
\phi_{in}}_{n}\!:)f^\sim_2(H,P).\eqno(5) $$
Here
$N(f_1,f_2)<\infty$ for functions $f_1,f_2$ with
compact support in momentum space.
 Thus,  the bilinear form (5) is a Wick polynomial and
it is easy to prove
that it defines a unique bounded operator.

Therefore, we can consider
the expressions
$$
(\Omega,\prod(f^\sim_j(H,P)\phi(0,0)f^\sim_{j+1}(H,P))\Omega).
$$
Due to translation invariance and the Schwartz nuclear theorem
 the expressions (6) allow to define
 Wightman functions
$ (\Omega,\prod\phi(t_j,x_j)\Omega) $
as
 generalized functions.
The direct consequences of
these considerations are the following statements:

\medskip
{\bf Theorem 2 (operator-valued generalized function).}
{\it The interacting field $\phi$ is a correctly
defined  operator-valued generalized function on the space
 $ {\cal F}(D(\R^4)) $
 in the Fock space of the free quantum $in$-field
 $\phi_{in} $. Here ${\cal F} $ the
 Fourier transform and $D(\R^4) $ is the Schwartz space
 of smoothed test functions
with compact support in momentum space.
 The analogous statement  is valid for the quantum outgoing
field $\phi_{out}. $
}
\medskip
\medskip

{\bf Theorem 3 (Wightman functions, \cite{Osi94b}).}
{\it The Wightman functions are correctly defined as
generalized functions on the space ${\cal F}(D(\R^{4n}))$.
They satisfy the positivity condition, the spectrum
 condition, $T$- and $P$-symmetry.
}

The analogous assertion is valid for
expressions
 $
(\Omega,\prod\phi_{\#}(t_j,x_j)\Omega) $
also.
Here
 $\phi_{\#}$ is either the $in$-field, or
the interacting field, or the $out$-field.
To construct the
quantum scattering operator (the $S$ matrix) and
to prove its unitarity
\cite{Osi96d,Jos65}
we need to consider these vacuum averages, too.

\medskip
\medskip
The next important properties of the quantum field  are
locality, unitarity
\cite{Osi96b,Osi96c}, and nontriviality
\cite{Osi94a,PedSZ92,Cal88}.

Here we shall formulate the assertion  about
 nontriviality
and we shall describe required steps for the proof
of locality and unitarity.

\medskip
\medskip
{\bf Theorem 4 (nontriviality,
\cite{Osi94a})}.
{\it The constructed $:\!\phi^4_4\!:$
 quantum field theory is nontrivial.
The coupling constant $\lambda$ is uniquely
defined by matrix elements of the interpolating quantum field.}

\medskip
\medskip
The consideration of locality
requires a further progress. An appropriate and important
step is the consideration of
a quantum field--\-bilinear form with
the help of
the Fock--\-Bargmann--\-Berezin--\-Segal integral
 and the integral representation
 in the $in$-field coordinates
\cite{Per86,PanPSZ91,BaeSZ92,Osi96c}.
This integral representation
 uses the Gaussian promeasure,
  Wick kernel (= Wick symbol), coherent states,
 holomorphic wave representation of the Fock space
and allows to write an integral representation for
singular operator and bilinear forms
\cite{PanPSZ91,Osi96c}.

We note, that in coordinates of the $in$-field
 the vacuum is described by Gaussian promeasure.
 The possibility to use coordinates
 of the $in$-field is connected
with the
 existence of wave operator.
 The existence of wave operator
simplifies consideration essentially.
A key moment in
a capability to apply  this integral
representation is the existence of the complex
structure (Theorem 1 and Theorem 2),
see
\cite{Osi96b,Osi96c,PanPSZ91}.

A bilinear form written with the help of
the Fock--\-Bargmann--\-Berezin--\-Segal
integral  representation
 has the following form
$$ \phi (\chi_1,\chi_2) =\int_ {H^ {1/2} (\R^3,\C)}
\int_ {H^ {1/2} (\R^3,\C)}
d\nu(z_1) d\nu (z_2) \phi (z_1, z_2)
\overline{\chi_1 (z_1)}
\chi_2 (z_2). \eqno (7)
$$
Here $\phi (z_1, z_2) $ is the Wick kernel
of the interacting quantum field,
$d\nu (z_1) $ is the promeasure corresponding
to the vacuum (in the $in$-field coordinates the promeasure
 is Gaussian),
 see \cite{PanPSZ91,Osi96c}.
 About promeasures, i.e. about
a consistent set of finite--\-dimensional
measures, or  about a close analogue of measures
on cylindrical sets, see, for instance,
\cite{PanPSZ91},
\cite[v. 4, ch. 4]{GelV61}.

This integral representation (7), estimates,
and the Bogoliubov ``edge of the wedge" theorem
\cite{Vla66} allow  to extend the  quantum field on
much more wide class  of test functions, including
functions  with compact support  in coordinate
space
\cite{Osi96c}.
This fact gives a possibility to consider locality.
In addition, locality requires the consideration of
 some
suitable finite--\-dimensional approximations and
the using of von Neumann's theorem about unitary equivalence
of finite-dimensional
 representations of algebra of canonical commutation relations
\cite{Osi96d}.

 Furthermore, the proof of  unitarity
is based on
 absence of bounded states (for the classical wave equation).

{\bf Conclusion.} The presented scheme of
construction of the $:\!\phi^4_4\!:$ quantum field theory
follows to the scheme, stated in
\cite{Hei74},
partially it is reflected
in
\cite{Osi84,Rac75}
(however, we note, that in
\cite{Rac75}
the iterations for the quantum field converge
and they allow to reconstruct the  quantum field,
but this reconstruction requires the existence of
   complex structure for solutions of the classical
nonlinear equation).
Long time a part of this scheme were
propagandized by Segal
(see, for example,
\cite{Seg63,Seg74}).
This scheme
is closely connected with the scheme of quantization
that have been proposed by Kostant
\cite{Kos70}.

It is interesting to compare the
stated exposition of the construction of the quantum
field in four-dimensional space-time
with the  known construction
in low dimension \cite{GliJ}.
We think that our scheme of exposition is
more appropriate
 from the physical point of view.
 It is interesting also
 to make comparison with the perturbation theory
 of the $:\!\phi^4_4\!:$ model.
 Note that the perturbation theory
 for the $:\!\phi^4_4\!:$ model is renormalizable .

There is wide  possibilities  to extend
 this scheme on higher dimensions,
on  higher degree of nonlinearities
(for which the Sobolev inequalities have the other form),
on Yang--\-Mills--\-Higgs fields
and similar questions.
A proper discussion of this issue is beyond the scope of this
Letter.

\medskip

This is the second announce paper  of the project
 $\phi^4_4 \cap M.$
The one of the goal of this project is to support
partly the Russian Fundamental Researches.

I acknowledge
A.~Kopilov, V.~Serbo,
 V.~Serebrjakov,
 members of Theoretical  Dept. of Budker Institute of Nuclear Physics,
 and
  Zinaida
    for the help and advice.

\end{document}